\documentclass[11pt]{article}
\usepackage[utf8]{inputenc}
\usepackage{abstract, amsmath, amssymb, graphicx, hyperref, adjustbox}
\usepackage[]{hyphenat}
\usepackage[normalem]{ulem}
\usepackage{natbib}
\usepackage{longtable}
\usepackage{booktabs}
\usepackage{lscape}
\usepackage{geometry}
\usepackage[graphicx]{realboxes}

\title{Optimizing Data-driven Weights In Multidimensional Indexes}
\author{
  Lidia Ceriani\thanks{University of Bologna, Department of Political and Social Sciences, \texttt{lidia.ceriani@unibo.it}} \and
  Chiara Gigliarano\thanks{Liuc University, School of Economics and Management, \texttt{cgigliarano@liuc.it}} \and
  Paolo Verme\thanks{University of Bologna, Department of Statistical Sciences ``Paolo Fortunati,'' \texttt{paolo.verme@unibo.it}} (Corresponding author)
}
\date{\today}

\begin{document}
\maketitle

\noindent
\rule{\textwidth}{.5pt}
\begin{abstract}
\noindent
Multidimensional indexes are ubiquitous, and popular, but present non-negligible normative choices when it comes to attributing weights to their dimensions. This paper provides a more rigorous approach to the choice of weights by defining a set of desirable properties that weighting models should meet. It shows that Bayesian Networks is the only model across statistical, econometric, and machine learning computational models that meets these properties. An example with EU-SILC data illustrates this new approach highlighting its potential for policies.\\


\end{abstract}
\rule{\textwidth}{.5pt}

\section{Introduction}
Social sciences, including economics and statistics, have recognized the importance of summary and synthetic measures that go beyond the representation of single variables or dimensions. Multidimensional indexes offer the opportunity to represent complex concepts such as well-being, happiness, or social capital with summary indicators. Examples of multidimensional indicators are Amartya Sen's indexes of well-being (\citealp{Sen1980}, \citealp{Sen1985}), the United Nations' Human Development Indexes, and the OECD's Better Life Index \citep{OECD2015}.

Despite their popularity, there is no consensus on how to aggregate dimensions and how weights should be attributed to each dimension. This has driven some organizations to adopt a ``dashboard'' approach where dimensions are monitored simultaneously, but separately. Examples of this approach are the OECD's Better Life Index \citep{OECD2015}, the Italian Equitable and Sustainable Well-being index \citep{BES2014}, and the United Kingdom's National Well-being Measure \citep{UK2015}. This approach highlights changes over time in each dimension, avoiding loss of information, but falls short of a parsimonious representation of well-being. 

A more popular approach is to use a single composite weighted index. Examples are the Human Development Index (HDI), the Canadian Index of well-being \citep{Canada2012}, the Happy Planet Index \citep{NEF2013}, and various multi-dimensional poverty or inequality indices used extensively by economists and statisticians (\citealp{Alkire2011}; \citealp{Aaberge2015}; \citealp{Bosmans2015}). The main drawbacks of this approach are the loss of information due to the extreme synthesis, and the arbitrariness of the choice of weights (particularly when weights are purely normative such as those based on researchers' or surveys respondents' opinions), and aggregation methods (\citealp{OECD2008}, \citealp{Decancq2013}, \citealp{Chiara2025}, \citealp{BELHADJ2012304}).

An additional shortcoming of multidimensional analyses shared by the dashboard and index methods is that dimensions are assumed to be independent (orthogonal). This is a rather unrealistic assumption as eloquently argued by the Fitoussi commission when discussing the quality of life index (\citealp{StiglitzSenFitoussi2010}): ``\emph{it is critical to address questions about how developments in one domain of quality of life affect other domains}'' (p.59) and ``\emph{when designing policies in specific fields, impacts on indicators pertaining to different quality-of-life dimensions should be considered jointly, to address the interactions between dimensions}'' (p.16). These statements highlight the importance of the complex correlation structure among dimensions and also the relevance of the direction of association between dimensions and the overall structure of causality.

This paper addresses these questions by arguing that data-driven weights in multidimensional indexes should be derived from models that meet a set of desirable properties designed to address the questions of orthogonality of dimensions, loss of information, and causality. After specifying these properties, the paper reviews statistical, econometric, and machine learning models to assess which model meets these properties. It finds that Bayesian Networks is the only model that satisfies all the desirable properties identified. An example shows how this model offers a practical approach to weighting and policy making with multidimensional indexes. 

\section{A Positive Data-driven Approach to Weighting}

When treating multidimensional concepts, scholars are faced with stark choices: ignoring the question of weights and monitor contributing dimensions separately (the ``dashboard'' approach), attribute equal weights, attribute different weights to different dimensions based on some normative criteria, or attribute weights based on some positive data-driven criteria. If one is searching for a single index with no priors on the importance of dimensions, the latter approach is understandably preferable, but there is no consensus on how to identify the optimal weighting model. In fact, there is hardly any discussion in the literature on what properties these data-driven models should meet. 

A possible and more rigorous approach is to state first the properties that models designed to study associations across variables should meet to be used for weighting dimensions in multidimensional indexes. Based on the conclusions of the Fitoussi commission, the shortcomings identified in the literature, and the existing features of popular models used in statistics, econometrics and machine learning, we argue that a plausible data-driven weighting model should meet the following criteria: 1) Measure the correlation between the outcome variable of interest and each of the dimensions included into the index; 2) Measure the correlation across the dimensions that constitutes the index; 3) Identify the direction of correlation among all variables and the structure of causality. We add that other desirable properties are 4) Being able to identify a latent variable when the outcome variable is not easily measured; 5) Be based on probabilistic rather than deterministic criteria to reduce the set of normative choices needed for identification; and 6) Be suitable for deriving relative weights from estimated coefficients. A model that satisfies these criteria, we argue, addresses the combination of issues that plague weighting in data-driven multidimensional indexes. 

The range of potential weighting models to consider is vast. In an effort to simplifying the analysis, and considering that multidimensional indexes typically provide a snapshot of the outcome considered, we excluded time models such as Granger, Markov, Panel, or Dynamic Bayesian Networks.\footnote{Note that longitudinal studies with multidimensional indexes exist (see for example \citealp{ZHANG2021109912}) but are few.} Otherwise, any model that studies cross-section associations among two or more variables can be considered a potential data-driven weighting model. It is important therefore to identify first a comprehensive list of models used across the social sciences to study association across variables, a rather ambitious task. 

To address this question, we sought the help of ChatGPT and Deepseek. We first compiled a list of popular models used to study association across sets of variables such as Spearman, OLS, Logit, Principal Component Analysis, Random Forest, and others. This initial list was forcibly limited by the authors' collective knowledge. We then asked both GPTs to complement this list with other popular models used across the social sciences. Next, we compared and assessed the lists provided, redacted a single list and submitted the list back to GPTs until authors and GPTs reached a consensus on a final list. We followed the same strategy to assess which model met which property by redacting an initial classification and ask both GPTs to comment until authors and GPTs reached an agreement on a final classification.

\begin{figure}[!htb]
\caption{Comparing Potential Weighting Models}\label{fig-plot}
\includegraphics[scale=0.6]{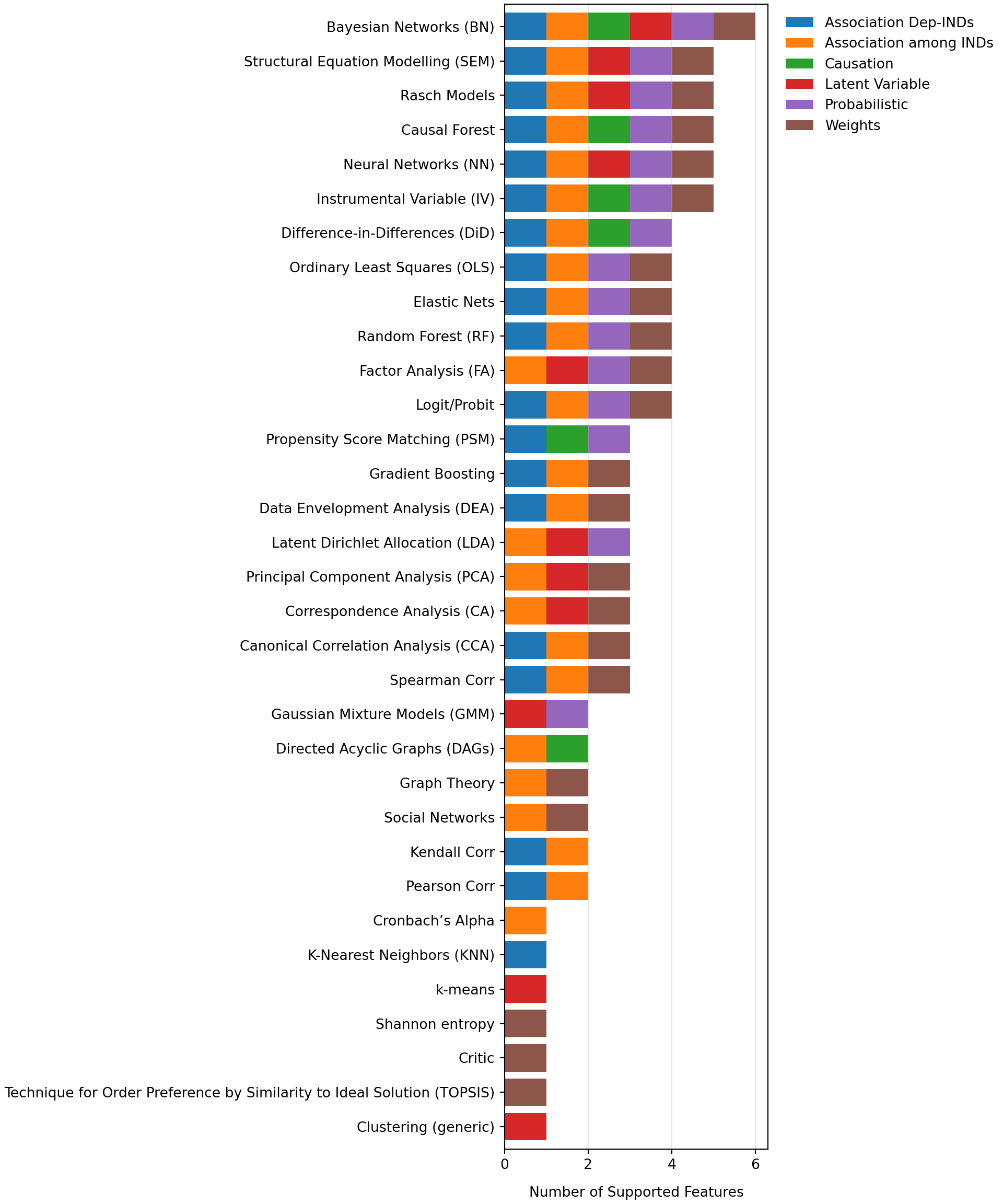}
\end{figure}

Figure \ref{fig-plot} summarizes results keeping models that met at least one of the desirable properties. The chart reports the number of properties that each model meets sorted in descending order. It shows that Bayesian Networks is the only model that satisfies all the six properties identified. Other models such as Structural Equation Modelling, Causal Forest and Neural Networks come close with five out of six properties satisfied whereas there are models that do not meet any property (not shown in chart), and others that meet only few. Also notable is the fact that some of the models that study associations well are not very suitable for constructing weights. Table A1 in Annex qualifies the findings and groups models into categories (Correlation, Regression, Machine Learning, Graphical Networks, Dimensionality Reduction, Latent Variable, Weighting, Diagnostic).\footnote{Note that several models would fit different groups. This is a coarse classification that is useful to see how classes of models behave. It is also useful to compare a reduced set of models that are representative of the main classes as shown further in the paper.} As shown by the Table, some of the findings may be disputable as several models meet some of the properties only under certain specifications. However, Bayesian Networks is the only model that consistently showed to satisfy all the desirable properties identified. Even if we removed the probabilistic condition, which understandably reduces the range of models that can be considered, Bayesian Networks would still sit on top of the classification. The only other model that can potentially meet all six conditions is the Structural Equation Model. However, SEM can be used for causal inference only under strict assumptions, and is computationally very challenging as compared to BNs.

\section{A Bayesian Networks Example}

\subsection{Data}

We provide an example of multidimensional index using an index of well-being based on the 2013 wave of EU-SILC, the European Union Statistics on Income and Living Conditions.  We consider \emph{life satisfaction} as our target (outcome) variable and include most of the key well-being dimensions suggested by the literature for these types of indexes including: \emph{material living standards}, (ii) \emph{health}, (iii) \emph{education}; (iv) \emph{personal activities and work}, (v) \emph{political voice and governance}, (vi) \emph{social connections and relationships}, (vii) \emph{security of physical as well as economic nature}, (iix) \emph{environment}, and also \emph{age}, \emph{gender} and \emph{country} as control variables. The unit of analysis is the household head. We include in the analysis all countries belonging to the European Union, with the exception of  Czech Republic, Denmark and Slovenia, because they lack information on one of the dimensions of social connections and relationships (variable pd050) (See \citealp{Liberati_2023} and \citealp{Bossert2013} for papers that use EU-SILC data for multidimensional indexes of well-being. See also \citealp{Ceriani2020} for a first application of Bayesian Networks to weighting in multidimensional indexes of well-being).

\subsection{A Bayesian Network of Dimensions}

We estimate a set of Bayesian Networks' structures using eleven different algorithms, including constraint-based learning algorithms, scored-based learning algorithms, and hybrid algorithms (we use the R package \textit{bnlearn} conceived by \citealp{Scutari2010}). We have also forced a set of arcs not to be included in the graph, by means of a blacklist, in order to avoid the control variables age and household size to be considered dependent of the well-being dimensions. Also, we have excluded the directed arcs from status in employment (WORK) and poverty (M\_POOR) to education (EDU) and from satisfaction with life (SA\_LIFE) to health (HEALTH). Moreover, we have whitelisted the directed arc from EDU to WORK, as our prior is that level of education should affect status in employment.

Table \ref{Tab_Learning_Algorithm} reports the arcs' occurrence in each of the  BN obtained by implementing the 11 algorithms. Following \citet{CugnataKenettSalini2016}, we assign a weight of 1 to arcs linking pairs of nodes directly and a weight of 0.5 if the connection is indirect. The last column in each table shows the sum of weights assigned to each pair of nodes, which can vary between 0 and 11. For the purpose of this paper, we define robust a BN which contains the largest set of arcs scoring 6 or more, meaning that the majority of algorithms find an occurrence.

\begin{figure}[!htb]
\caption{Robust BN (Tabu-aic algoritm)}\label{Figure2}
\includegraphics[scale=0.8]{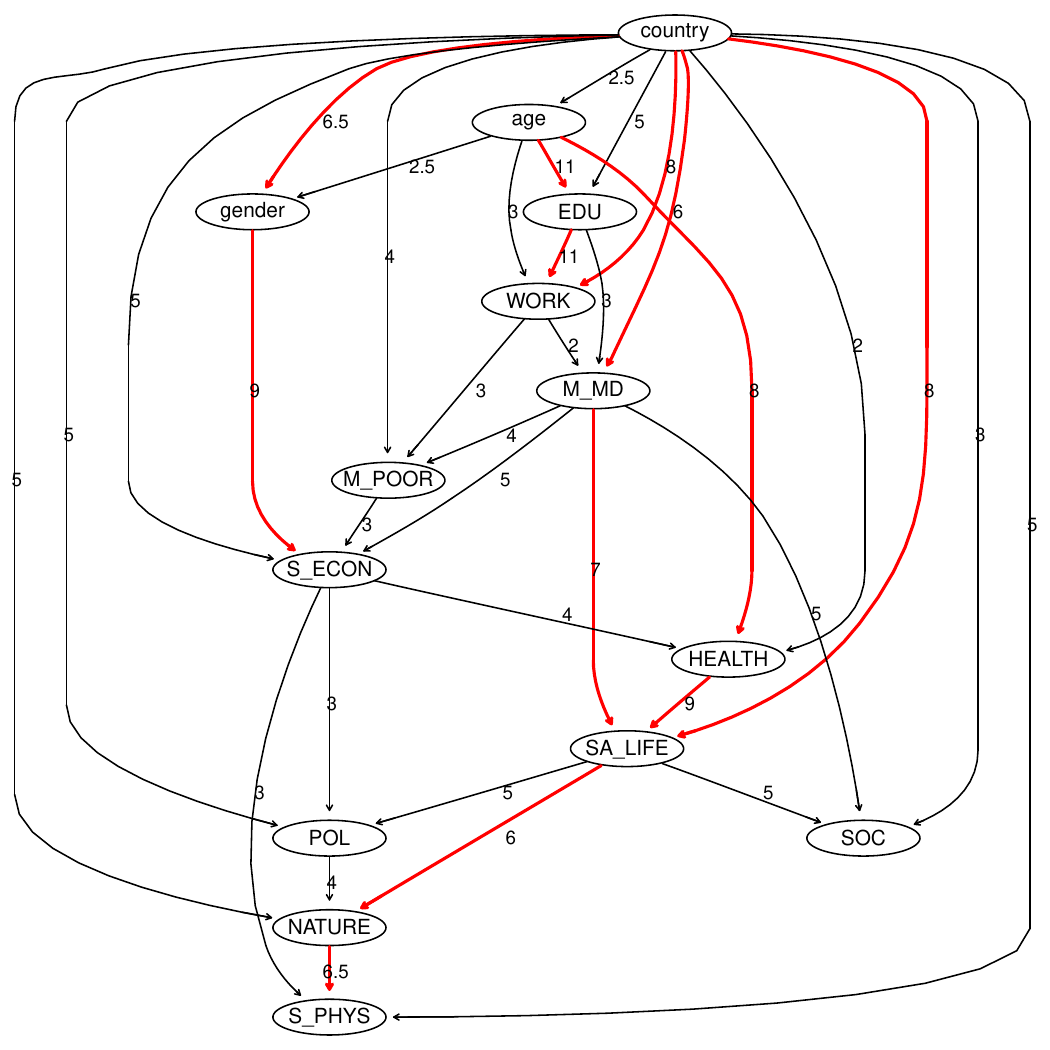}
\end{figure}

The network resulting from applying the Tabu-aic algorithm in Figure \ref{Figure2} contains the highest number of arcs among all networks produced by the 11 different algorithms applied. The numbers shown on the arcs represent the number of algorithms confirming the directional connections. See, for example, age to EDU and EDU to WORK, which correspond to the first two rows of Table \ref{Tab_Learning_Algorithm}. The arcs that appear in at least 6 of the 11 algorithms used in the analysis are depicted in red. Variables of interest (non-control variables) are in capital letters. The numbers on the arcs represent the number of algorithms confirming the directional connection. In red the arcs contained in more than 5 algorithms.

Of the different well-being dimensions taken into account, only education (EDU), personal activities and work (WORK), material living standards (M\_MD and  M\_\allowbreak POOR), economic security (S\_ECON) and health status (HEALTH) influence the target variable (satisfaction with life, SA\_LIFE). On the other hand, political and social participation (SOC and POL), satisfaction with the environment (NATURE) and physical security (S\_PHYS) appear to be consequences of the level of life satisfaction (SA\_LIFE). These findings are sensible but recall that researchers may whitelist or blacklist variables based on prior knowledge or common sense. In other words, researchers can impose as much structure as they like on the BN following the initial data-driven analysis.

\subsection{Generating Weights From Bayesian Networks}

The information on the strength of the relationship between nodes can be naturally translated in the weighting structure of a multidimensional index of well-being. In this new system of weights, dimensions that have a larger impact on the target node  when affected by a policy-induced change (e.g. have \emph{more influence} on the target node) receive a higher weight.

Let us assume that $T$ is the target node (\emph{satisfaction with life}). For each $i-$th dimension of well-being, the weight assigned to dimension $X_i$ is positively correlated with the strength of the arcs connecting $X_i$ and $T$ and negatively correlated to the length of the path linking $X_i$ with $T$. Formally:

\begin{equation}\label{eq_weight}
w_i(X_i, T; p_i) = \sum_{p_i\in P_i(X_i, T)} \sigma_i^{|p_i|}
\end{equation}

where $P_i(X_i, T)$ is the set of paths in the BN joining nodes $X_i$ and $T$, $|p_i|$ is the length  of each path and $\sigma_i$ is the strength associated to each path. A path is defined as the set of arcs connecting each node $X_i$ to the target node $T$. The strength of each path corresponds to the product of the strengths of each arc in the path (which are bounded between 0--no influence between two nodes and  1--maximum influence). This definition of weight corresponds to the \emph{distance-weighted influence} in \citet{Albrecht2014} and \citet{CugnataKenettSalini2016}.

The proposed definition of weights represents an improvement on the classical \emph{data-driven} and \emph{normative} approaches to weights found in the literature (see \citealp{Decancq2013} for a discussion). As for \emph{data-driven} weights, BN weights consider the distribution of achievements in society and, as in the \emph{normative} approach, they attribute more weight to policy relevant variables. Unlike these approaches, BN weights include data-driven information on inter-dependencies among variables and can be used for \textit{ex-ante} prognostic (forecasting) and \textit{ex-post} diagnostic (evaluation) policy purposes. By providing this extra information, they also partly address the criticism of loss of information attributed to the multidimensional index approach.

\section{Comparing Different Weighting Schemes}

As a further illustration, we compare the resulting rankings of selected weighting schemes representing the main classes of models including: (i) Spearman Correlation; (ii) Linear Regression; (iii) Random Forest; and iv) Bayesian network. As benchmarks, we also use v) Equal Weighting; and vi) Self-reported opinions from Eurobarometer. Such comparison is parsimonious enough to allow for a visual assessment while comparing different classes of weighting schemes. 

As multivariate correlation-based method, we use the Spearman rank correlation coefficient, which seems the most appropriate given the categorical nature of our underlying data (see \citealp{Banerjee_2018} for a discussion on deriving multidimensional weights using multidimensional coefficient of variation). Each dimension weight is defined as the absolute value of the correlations coefficient ($\rho$) between the dimension and the target variable $t$, plus half the value of each other pairwise $\rho$'s:

\begin{equation*}
w^{SC}_i = \|\rho_{i,t}\| + \frac{1}{2}\sum_{j\neq i,t} \|\rho_{i,j}\|
\end{equation*}

Weights are then normalized dividing by $\sum_{i \neq t} w_i$. Notably, this method offers an improvement over linear estimation by capturing pairwise correlations among independent variables; however, it remains limited in its ability to reveal the broader network of interrelationships across dimensions.

As machine learning method, we apply Random Forest regression algorithms, and we use as weights the predicted importance associated to each independent variable, $\iota_i$, normalized by the sum of all importance values associated to each dimension $i-$th, $i=1, \dots, m$:

\begin{equation*}
w^{RF}_i = \frac{\iota_i}{\sum_{i=1}^m \iota_i}
\end{equation*}

For the opinion-based weights, we use data from the Eurobarometer survey, wave 86.3 fielded between November and December 2016. The weight are represented by the share of the population expressing strong concern with respect to selected issues, following \citet{Guio2009} and \citet{Bossert2013}. Table \ref{tab:weights} summarizes the variables used to capture the relative importance that individuals attach to each dimension.

The application of multiple weighting methods reveals substantial variation in how well-being dimensions are prioritized (see Table \ref{tab:weights}). From a policy perspective, the equal weighting approach offers limited analytically utility, as it fails to account for the varying influence of each dimension. Subjective weights highlight Satisfaction with One’s Economic Situation, Social Participation, Health, and Satisfaction with Work as the most salient contributors to overall life satisfaction, reflecting individual or cultural perceptions of well-being. In contrast, data-driven methods such as OLS regression, Random Forest, and Spearman correlation produce more balanced distributions, yet consistently assign greater importance to Health, Material Deprivation, and Economic Satisfaction. Among these, Random Forest produces a particularly even distribution of weights, assigning relatively lower importance to Poverty Status, Economic Satisfaction, and Physical Security, while giving slightly more weight to Material Deprivation, Environmental Satisfaction, and Satisfaction with Work.

Bayesian Networks, on the other hand, allocate nearly all the weight to Health and Material Deprivation. This concentration reflects the model’s ability to account for both direct and indirect effects, revealing that these two dimensions exert the greatest overall influence on life satisfaction. Their centrality arises not only from their direct impact but also from their role in shaping other dimensions within the well-being framework. This capacity to capture indirect pathways is a key advantage of the Bayesian Network approach over alternative methods. Specifically, it offers three critical benefits: (i) it incorporates indirect effects, providing a more comprehensive understanding of causal influence; (ii) it narrows the range of policy actions required to improve outcomes; and (iii) it enhances the cost-effectiveness of interventions by leveraging the amplifying potential of indirect relationships.

Figure 3 clearly demonstrates that the choice of weighting method is not neutral: it can significantly shift a country's rank in a composite well-being index. While data-\allowbreak driven methods tend to offer more consistent patterns, subjective or structure-\allowbreak based approaches (like Eurobarometer and Bayesian Networks) can produce markedly different outcomes, especially for countries in the middle of the distribution. In particular, the large changes observed in the ranks of Malta (MT), Latvia (LV), and Lithuania (LT) are due to the BN method placing almost all the weight on the HEALTH and M\_MD dimensions. In these dimensions, LV and LT rank among the worst performers, while MT ranks among the best across all countries considered. Conversely, in the remaining dimensions, LV and LT are among the best performers, whereas MT ranks among the worst. With all other methods, the relative performance across all dimensions is smoothed out, resulting in more moderate changes in country rankings.

\section{Conclusions}

The paper showed that Bayesian Networks is the only model meeting a set of desirable properties identified for weighting in multidimensional indexes. This model offers a clear graphical representation of the hierarchical sets of relations across dimensions, it can be easily used to derive weights, and results in clearer, more parsimonious, and potential cost-efficient policy indications.

\newpage
\bibliographystyle{chicago}
\bibliography{References}

\newpage
\centering

\newpage
\centering

\newpage
\centering
\begin{figure}[!htb]
\caption{Ranking of countries according to different weighting scheme}\label{Figure3}
\includegraphics[scale=0.35]{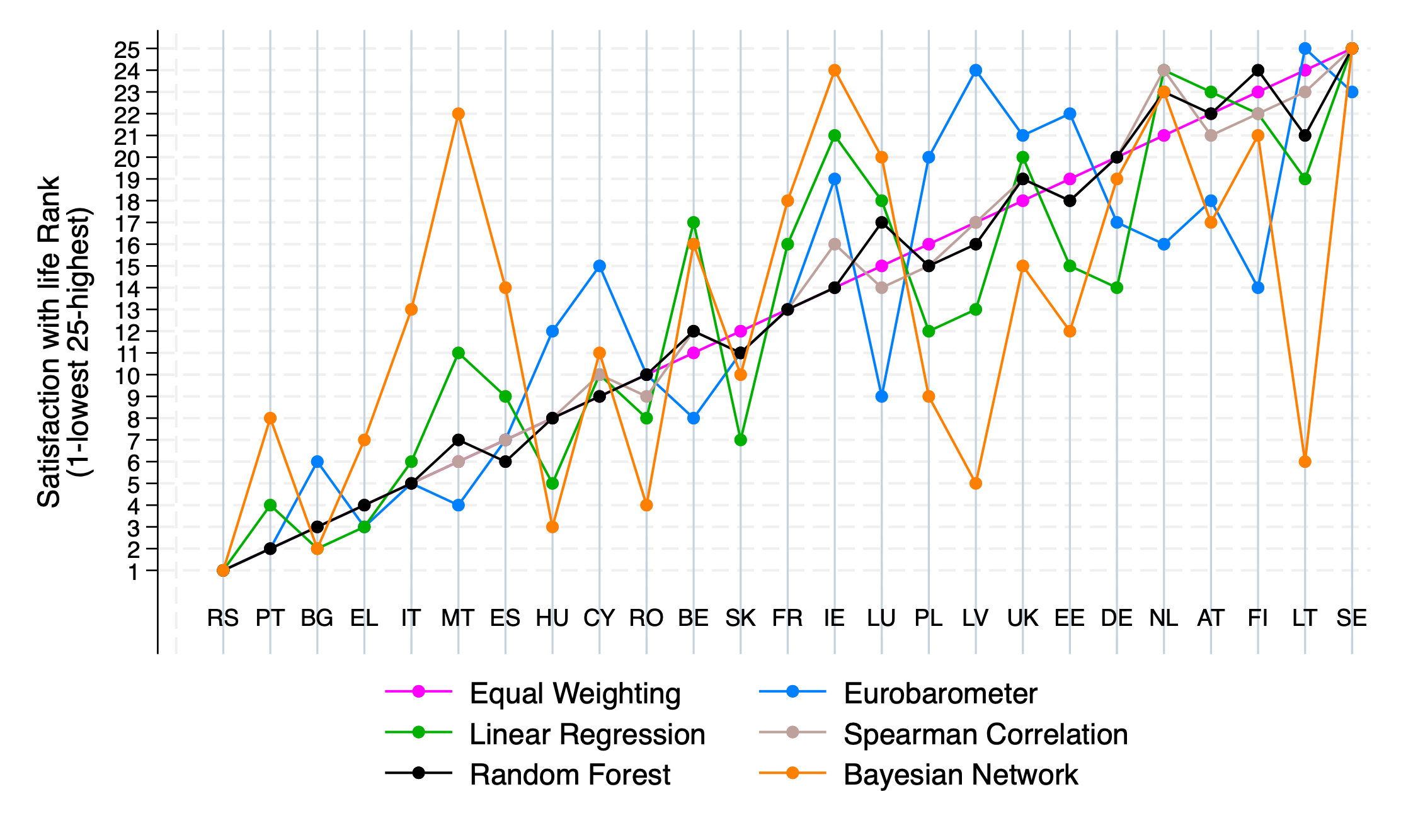}
\end{figure}

\newpage

\begin{table}[!ht]
\small
\caption{List of arcs from different learning algorithms}\label{Tab_Learning_Algorithm}
\begin{tabular}{llccccc|cccc|cc|c}
\hline
& & \multicolumn{5}{c|}{Score-based} & \multicolumn{4}{c|}{Constraint-based} & \multicolumn{2}{c|}{Hybrid} & TOT\\
\multicolumn{2}{c}{Arcs}&	 \multicolumn{3}{c}{HC}	&	\multicolumn{2}{c|}{TABU}	&	GS	&	\multicolumn{3}{c|}{INCAS}	&	\multicolumn{2}{c|}{MMHC}	&		\\
\multicolumn{2}{c}{}&	 bic	&	aic	&	k2	&	bic	&	aic	&	&		&	fast	&	int	&	bic	& rm &		\\
\hline
age	&	EDU	&	1	&	1	&	1	&	1	&	1	&	1	&	1	&	1	&	1	&	1	&	1	&	11	\\
EDU	&	WORK	&	1	&	1	&	1	&	1	&	1	&	1	&	1	&	1	&	1	&	1	&	1	&	11	\\
HEALTH	&	SA\_LIFE	&	1	&	1	&	1	&	1	&	1	&	0	&	1	&	0	&	1	&	1	&	1	&	9	\\
gender	&	S\_ECON	&	0	&	1	&	1	&	0	&	1	&	1	&	1	&	1	&	1	&	1	&	1	&	9	\\
age	&	HEALTH	&	1	&	1	&	1	&	1	&	1	&	0	&	1	&	0	&	1	&	1	&	0	&	8	\\
country	&	WORK	&	1	&	1	&	1	&	1	&	1	&	1	&	0	&	0	&	1	&	0	&	1	&	8	\\
country	&	SA\_LIFE	&	0	&	1	&	1	&	0	&	1	&	1	&	1	&	1	&	1	&	1	&	0	&	8	\\
M\_MD	&	SA\_LIFE	&	1	&	1	&	1	&	1	&	1	&	0	&	1	&	0	&	1	&	0	&	0	&	7	\\
country	&	gender	&	1	&	1	&	1	&	1	&	1	&	0.5	&	0.5	&	0.5	&	0	&	0	&	0	&	6.5	\\
NATURE	&	S\_PHYS	&	1	&	1	&	1	&	1	&	1	&	0	&	0	&	0.5	&	0	&	1	&	0	&	6.5	\\
country	&	M\_MD	&	1	&	1	&	1	&	1	&	1	&	0	&	0	&	0	&	1	&	0	&	0	&	6	\\
SA\_LIFE	&	NATURE	&	1	&	1	&	1	&	1	&	1	&	0	&	0	&	0	&	0	&	1	&	0	&	6	\\
country	&	NATURE	&	1	&	1	&	1	&	1	&	1	&	0	&	0	&	0	&	0	&	0	&	0	&	5	\\
country	&	POL	&	1	&	1	&	1	&	1	&	1	&	0	&	0	&	0	&	0	&	0	&	0	&	5	\\
country	&	S\_ECON	&	1	&	1	&	1	&	1	&	1	&	0	&	0	&	0	&	0	&	0	&	0	&	5	\\
country	&	S\_PHYS	&	1	&	1	&	1	&	1	&	1	&	0	&	0	&	0	&	0	&	0	&	0	&	5	\\
M\_MD	&	S\_ECON	&	1	&	1	&	1	&	1	&	1	&	0	&	0	&	0	&	0	&	0	&	0	&	5	\\
M\_MD	&	SOC	&	1	&	1	&	1	&	1	&	1	&	0	&	0	&	0	&	0	&	0	&	0	&	5	\\
SA\_LIFE	&	POL	&	1	&	1	&	1	&	1	&	1	&	0	&	0	&	0	&	0	&	0	&	0	&	5	\\
SA\_LIFE	&	SOC	&	1	&	1	&	1	&	1	&	1	&	0	&	0	&	0	&	0	&	0	&	0	&	5	\\
country	&	EDU	&	0	&	1	&	1	&	0	&	1	&	0	&	1	&	0	&	1	&	0	&	0	&	5	\\
M\_MD	&	M\_POOR	&	1	&	1	&	0	&	1	&	1	&	0	&	0	&	0	&	0	&	0	&	0	&	4	\\
S\_ECON	&	HEALTH	&	1	&	1	&	0	&	1	&	1	&	0	&	0	&	0	&	0	&	0	&	0	&	4	\\
country	&	M\_POOR	&	0	&	1	&	1	&	0	&	1	&	1	&	0	&	0	&	0	&	0	&	0	&	4	\\
POL	&	NATURE	&	0	&	1	&	0	&	0	&	1	&	0	&	1	&	0	&	1	&	0	&	0	&	4	\\
S\_PHYS	&	NATURE	&	0	&	0	&	0	&	0	&	0	&	0	&	1	&	0.5	&	1	&	0	&	1	&	3.5	\\
$\dots$ & $\dots$ & $..$ & $..$ & $..$ & $..$ & $..$ & $..$ & $..$ & $..$ & $..$ & $..$ & $..$ & $..$\\
\hline
\multicolumn{14}{l}{Note: Arcs supported by less than 3 algorithms are not reported.}
\end{tabular}
\end{table}  

\newpage

\begin{table}[!ht]
\centering
\caption{Weights (in \%) assigned to wellbeing dimensions using different methods} \label{tab:weights}
\begin{small}
\begin{tabular}{lrrrrrr}
\hline
Dimension & EQ  & EB & RE & SP & RF & BN \\
\hline
Education (EDU)&10&6.6&.7&10&9.6&0 \tabularnewline
Health (HEALTH)&10&11.2&21.5&10.4&11&46.9 \tabularnewline
Material Deprivation (M\_MD)&10&9.9&12.8&11.6&15.9&52.7 \tabularnewline
Being Poor (M\_POOR)&10&8.8&9.5&8.1&6.9&0 \tabularnewline
Satisfaction with Nature (NATURE) &10&6.8&10.6&11.7&9.7&0 \tabularnewline
Political Participation (POL)&10&7.5&11.7&10.4&10.2&0 \tabularnewline
Social Participation (SOC) &10&12.3&16.8&10.5&10.9&0 \tabularnewline
Economic Security (S\_ECON)&10&21.1&12.6&8.6&13.2&.4 \tabularnewline
Physical Security (S\_PHYS)&10&4.1&2.4&8.1&3&0 \tabularnewline
Work Satisfaction (WORK)&10&11.9&1.4&10.6&9.7&0 \tabularnewline
\hline
\end{tabular}

Source: Authors' calculations. Bayesian Networks (BN), Subjective Weights (EB), Equal Weights (EQ), OLS Regression (RE), Random Forest (RF), and Spearman Correlation (SP).
\end{small}
\end{table}

\newpage
\section*{Appendix 1}
\setcounter{table}{0}
\renewcommand{\thetable}{A\arabic{table}}
\begin{table}[!htb]
\caption{Comparing Potential Weighting Models}
\begin{scriptsize}
\begin{adjustbox}{angle=90}
\begin{tabular}{llllll}
	&	Group	&	Models	&	Association Dep-INDs	&	Association among INDs	&	Causation	\\
\hline											
1	&	Corr	&	Pearson	&	Yes	&	Yes (bivariate)	&	No	\\
2	&	Corr	&	Spearman	&	Yes	&	Yes (bivariate)	&	No	\\
3	&	Corr	&	Kendall	&	Yes	&	Yes (bivariate)	&	No	\\
4	&	Corr	&	CCA	&	Yes	&	Yes (multivariate sets)	&	No	\\
5	&	Reg	&	OLS	&	Yes	&	Yes (DV vs. IVs)	&	No (unless assumptions hold)	\\
6	&	Reg	&	Logit/Probit	&	Yes	&	Yes (DV vs. IVs)	&	No (unless causal design)	\\
7	&	Reg	&	IV	&	Yes	&	Yes (DV vs. IVs, accounting for endog.)	&	Yes (with valid IV)	\\
8	&	Reg	&	DiD	&	Yes	&	Yes (causal DV-IV)	&	Yes (under parallel trends)	\\
9	&	Reg	&	EN	&	Yes	&	Yes (DV vs. IVs)	&	No	\\
10	&	Reg	&	SEM	&	Yes	&	Yes (network of DV-IV links)	&	Yes (if causal model specified)	\\
11	&	Reg	&	PSM	&	Yes (causal)	&	No (balances groups, no direct assoc.)	&	Yes	\\
12	&	ML	&	RF	&	Yes	&	Yes (implicit feature interactions)	&	No	\\
13	&	ML	&	Causal Forest	&	Yes	&	Yes (causal DV-IV)	&	Yes (heterogeneous effects)	\\
14	&	ML	&	Gradient Boosting	&	Yes	&	Yes (implicit feature interactions)	&	No	\\
15	&	ML	&	NN	&	Yes	&	Yes (complex nonlinear assoc.)	&	No	\\
16	&	ML	&	KNN	&	Yes (predictive)	&	No (instance-based, no formal assoc.)	&	No	\\
17	&	ML	&	Clustering (generic)	&	No (unsupervised)	&	No (groups observations)	&	No	\\
18	&	ML	&	k-means	&	No (unsupervised)	&	No (groups observations)	&	No	\\
19	&	ML	&	GMM	&	No (unsupervised)	&	No (groups observations)	&	No	\\
20	&	ML	&	LDA	&	No (topic modeling)	&	Yes (doc-word co-occurrence)	&	No	\\
21	&	GN	&	BN	&	Yes (probabilistic)	&	Yes (conditional dependencies)	&	Yes (if causal structure known)	\\
22	&	GN	&	DAGs	&	No (framework)	&	Yes (represents causal assoc.)	&	Yes (theoretical basis)	\\
23	&	GN	&	Graph Theory	&	No (math framework)	&	Yes (abstract relational structure)	&	No	\\
24	&	GN	&	Social Networks	&	No (relational analysis)	&	Yes (node-edge relationships)	&	No (unless causal design)	\\
25	&	DR	&	PCA	&	No (dim. reduction)	&	Yes (covariance among variables)	&	No	\\
26	&	DR	&	FA	&	No (latent structure)	&	Yes (latent variable assoc.)	&	No	\\
27	&	LV	&	CA	&	No	&	Yes	&	No	\\
28	&	LV	&	Rasch Models	&	Yes	&	Yes (item-person links)	&	No	\\
29	&	WE	&	Shannon entropy	&	No (info theory)	&	No (measures uncertainty)	&	No	\\
30	&	WE	&	Critic	&	No (weighting method)	&	No (objective weights)	&	No	\\
31	&	WE	&	TOPSIS	&	No (MCDA method)	&	No (ranking alternatives)	&	No	\\
32	&	WE	&	DEA	&	Yes (input-output)	&	Yes (efficiency frontiers)	&	No	\\
33	&	DG	&	VIF	&	No (diagnostic tool)	&	No (measures multicollinearity)	&	No	\\
34	&	DG	&	Cronbach’s Alpha	&	No	&	Yes	&	No	\\
\hline											
\end{tabular}
\end{adjustbox}
\end{scriptsize}
\end{table}

\newpage
\begin{table}[!htb]
\begin{scriptsize}
\begin{adjustbox}{angle=90}
\begin{tabular}{lllllll}
	&	Group	&	Models	&	Latent Variable	&	Probabilistic	&	Weights	&	Reference	\\
\hline													
1	&	Corr	&	Pearson	&	No	&	No	&	No	&	OECD (2008)	\\
2	&	Corr	&	Spearman	&	No	&	No	&	Yes (only for diagnostic)	&	Mazziotta and Pareto (2016)	\\
3	&	Corr	&	Kendall	&	No	&	No	&	No	&	Not Found	\\
4	&	Corr	&	CCA	&	No	&	No	&	Yes (canonical weights)	&	Härdle and Simar (2015)	\\
5	&	Reg	&	OLS	&	No	&	Yes (with inference)	&	Yes (coefficients as weights)	&	Cherchye et al. (2007)	\\
6	&	Reg	&	Logit/Probit	&	No	&	Yes	&	Yes (coefficients as weights)	&	Krishnakumar (2007)	\\
7	&	Reg	&	IV	&	No	&	Yes	&	Yes (coefficients as weights)	&	Heckman and Vytlacil (2007)	\\
8	&	Reg	&	DiD	&	No	&	Yes (with SEs)	&	No	&	Not Found	\\
9	&	Reg	&	EN	&	No	&	No (unless Bayesian)	&	Yes (shrunk coefficients)	&	Greco et al. (2019)	\\
10	&	Reg	&	SEM	&	Yes (common in SEM)	&	Yes	&	Yes (path coefficients)	&	Booysen (2002)	\\
11	&	Reg	&	PSM	&	No	&	Yes (with uncertainty)	&	No	&	Not Found	\\
12	&	ML	&	RF	&	No	&	No (unless Bayesian RF)	&	Yes (feature importance)	&	Berrar et al. (2016)	\\
13	&	ML	&	Causal Forest	&	No	&	Yes	&	Yes (feature importance)	&	Athey and Imbens (2016)	\\
14	&	ML	&	Gradient Boosting	&	No	&	No	&	Yes (feature importance)	&	Chen and Guestrin (2016)	\\
15	&	ML	&	NN	&	Yes (context-dependent)	&	Yes (if Bayesian)	&	Yes (attention weights)	&	Kokotović et al. (2020)	\\
16	&	ML	&	KNN	&	No	&	No	&	No	&	Not Found	\\
17	&	ML	&	Clustering (generic)	&	Yes (latent clusters)	&	No	&	No	&	OECD (2008)	\\
18	&	ML	&	k-means	&	Yes (latent clusters)	&	No	&	No	&	Asselin (2009)	\\
19	&	ML	&	GMM	&	Yes (latent distributions)	&	Yes	&	No	&	Bartholomew et al. (2008)	\\
20	&	ML	&	LDA	&	Yes (latent topics)	&	Yes	&	No	&	Not Found	\\
21	&	GN	&	BN	&	Yes (latent nodes possible)	&	Yes	&	Yes (conditional probabilities)	&	Delgado et al. (2018)	\\
22	&	GN	&	DAGs	&	No	&	No	&	No	&	Not Found	\\
23	&	GN	&	Graph Theory	&	No	&	No	&	Yes	&	Cobo et al. (2011)	\\
24	&	GN	&	Social Networks	&	No	&	No	&	Yes	&	Freeman (1979)	\\
25	&	DR	&	PCA	&	Yes (latent components)	&	No	&	Yes (loadings as weights)	&	Jolliffe (2002)	\\
26	&	DR	&	FA	&	Yes (primary purpose)	&	Yes	&	Yes (factor loadings)	&	OECD (2008)	\\
27	&	LV	&	CA	&	Yes	&	No	&	Yes	&	Greenacre (1984)	\\
28	&	LV	&	Rasch Models	&	Yes	&	Yes	&	Yes (item difficulty params)	&	Fusco and Dickens (2008)	\\
29	&	WE	&	Shannon entropy	&	No	&	No	&	Yes	&	Zeleny (1982)	\\
30	&	WE	&	Critic	&	No	&	No	&	Yes (explicit weighting method)	&	Diakoulaki et al. (1995)	\\
31	&	WE	&	TOPSIS	&	No	&	No	&	Yes (explicit weighting method)	&	Hwang and Yoon (1981)	\\
32	&	WE	&	DEA	&	No	&	No	&	Yes (custom DMU weights)	&	Cherchye et al. (2007)	\\
33	&	DG	&	VIF	&	No	&	No	&	No	&	Not Found	\\
34	&	DG	&	Cronbach’s Alpha	&	No	&	No	&	No	&	OECD (2008)	\\
\hline	
\multicolumn{7}{l}{
\begin{minipage}{21cm}
Legenda: Corr=Correlation; Reg=Regression; ML=Machine Learning;
BN=Bayesian Networks; GN=Graphical Networks; DR=Dimensionality
reduction; LV=Latent Variable; WE=Weighting; DG=Diagnostic; Canonical
Correlation Analysis (CCA), Ordinary Least Squares (OLS), Instrumental
Variable (IV), Difference-in-Differences (DiD), Elastic Nets (EN),
Structural Equation Modeling (SEM), Propensity Score Matching (PSM),
Random Forest (RF), Neural Networks (NN), K-Nearest Neighbors (KNN),
Gaussian Mixture Models (GMM), Latent Dirichlet Allocation (LDA),
Bayesian Networks (BN), Directed Acyclic Graphs (DAGs), Principal
Component Analysis (PCA), Factor Analysis (FA), Correspondence Analysis
(CA), Technique for Order Preference by Similarity to Ideal Solution
(TOPSIS), Data Envelopment Analysis (DEA), and Variance Inflation Factor
(VIF)
\end{minipage}}
\end{tabular}												
\end{adjustbox}
\end{scriptsize}
\end{table}

\end{document}